\documentclass[12pt,a4paper]{article}

\usepackage{amsmath}

\usepackage{times}
\usepackage{mathptmx}
\usepackage{epic}
\usepackage{epsfig}

\makeatletter
\def\NAT@parse{\typeout{Error: Attempt to use fake Natbib command 
which is provided to fool Hyperref.}}
\makeatother
\usepackage{cite}

\title{Multi-Agent Model Predictive Control:\\A Survey}
\author{R.R.~Negenborn$^{\footnotesize *}$, B.~De Schutter, J.~Hellendoorn\\
\normalsize Delft Center for Systems and Control,\\\normalsize Delft University of Technology\\\normalsize Mekelweg 2, 2628 CD Delft, The Netherlands\\\small * corresponding author, e-mail: r.negenborn@dcsc.tudelft.nl}
\date{Technical report 04-010\\\ \\\normalsize\today}

\newcommand{\etal}{\emph{et al.}}

\begin{document}

\maketitle

\begin{abstract}
\noindent In this report we define characteristic control design elements
and show how conventional single-agent MPC implements these. We survey recent
literature on multi-agent MPC and discuss how this literature deals with
 decomposition, problem assignment, and cooperation.
\end{abstract}


\section{Introduction}

Already back in 1978, Sandell \etal
\cite{San:78} surveyed a wide range of alternative methods for decentralized
control. They find that a good combination of engineering judgment and
analysis can be used to define in a reasonable way an ad-hoc control structure
for a dynamic system. They conclude that methodologies are needed that
present a designer with several good control structure candidates for further
consideration.

In this report we look at how research since 1978 has advanced
distributed control. We consider the control of large-scale systems like power
networks, traffic networks, digital communication networks, flexible
manufacturing networks, ecological systems, etc. In particular, we survey some
of the literature on Model Predictive Control (MPC) in distributed settings. We will
refer to this as \emph{Multi-Agent Model Predictive Control}. We are
interested in the control design methods that have been developed so far.

The structure of this reported is as follows. In order to classify and find structure in the literature on multi-agent MPC,
in Section \ref{sec:cd} we first consider control methodologies in
general. Control methodologies involve different kinds of models. Depending on
the actual models chosen, different issues rise that have to be considered.
In Section \ref{sec:sampc} we focus on Model Predictive Control (MPC). We
explain the general idea behind MPC and characterize the MPC framework in terms
of the models of Section \ref{sec:cd}. As it turns out, the standard MPC
framework may be seen as single-agent MPC.
In Section \ref{sec:mampc} we move on to the discussion of multi-agent MPC. We refer to multi-agent MPC as a
general term for methods that apply the MPC strategy using multiple agents to
control a system. Important aspects of multi-agent MPC are the way in which a
system is decomposed into subsystems (centralized, decentralized, hierarchical),
the way in which control problems are formulated on these decomposed systems
 (centralized, decentralized, hierarchical),
and the way in which agents communicate with one another in order to solve these
control
problems. We describe how recent literature on multi-agent MPC implements these
issues.
Finally,  we end this report with open issues and concluding remarks in Section
\ref{sec:conc}.


\section{Control Methodologies}\label{sec:cd}

In this section we consider different types of concepts that play a role in general control methodologies. We consider the underlying task of control problems, system models that may be used for control, control problem models formulating a control problem, and agent architectures useful in solving control problems.

\subsection{Control Task}
\begin{figure}
\centering \includegraphics[width=0.6\textwidth]{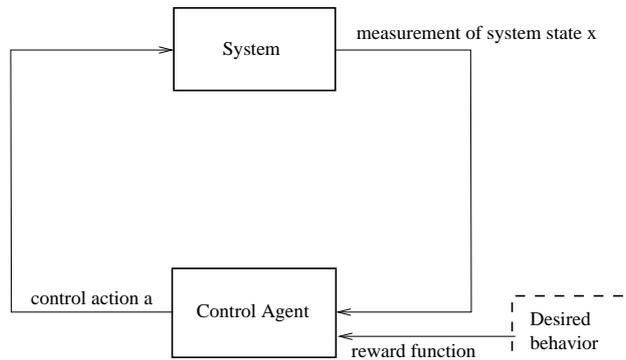}
\caption{General scheme for controlling a system. The control agent measures the state of the system and determines an action such that the behavior of the system approaches the desired behavior as close as possible.
}
\label{fig:control}
\end{figure}
In a control context, typically a system is supposed to behave in a certain way.
It should accomplish some task, which may involve reaching a certain number of goals. The task has to be accomplished while making sure that any possible constraints are not violated.

Tasks may be provided by a human or some
artificial entity, or they may follow from some behavioral characteristics or
reasoning of the system. Goals can be short-term goals, e.g., to bring the system in a certain state, or long-term, e.g., to maximize the long-term performance or to minimize the long-term operation costs. Note that tasks need not have one single goal. They may have multiple, possibly conflicting, goals. In that case they are referred to as so-called
\emph{multi-objective} tasks.

Actions that can be performed on the system have to be chosen in such a
way that the task of the system is achieved, keeping in mind the dynamics of the
system, and possible constraints on the actions. Finding the actions that
achieve the goal is called the \emph{Dynamic Control Problem} (DCP). A typical DCP setting is shown in Figure \ref{fig:control}. The general DCP can be formulated as:
\begin{quote}
\textit{Find the actions such that the goal is achieved optimally}\\
\\
\textit{subject to}
\begin{quote}
\textit{a model of the dynamic system}\\
\textit{including constraints on the actions and states.}
\end{quote}
\end{quote}
This problem can be seen as an \emph{optimization problem}, since the actions
have to be chosen such that they achieve the goal in the best possible way. Note
that the goal is independent of the model of the dynamic system.

This section defines the elements that play a role in controlling a general
system. As mentioned, controlling a system comes forth from having the desire to
have the system achieve a certain goal. In order to obtain the goal we will
assume that there is a \emph{system model} of the system under
consideration.\footnote{It is not always necessary to have a model of the
system, e.g. when using PID controllers.} Such a model can be used to define
more precisely what the goal is and to predict how the system will behave given
certain actions. A \emph{control problem model} defines what the
exact control problem is, often based on the system model. A control problem model is
used by agents that solve the problem. The agents are organized in an
\emph{agent architecture} and follow some \emph{communication protocol}.

\subsection{System Models}\label{sec:dynmod}

Dynamic \emph{system models} describe the
behavior of the physical system given \emph{actions} on the system, the
\emph{state} of the system, and possibly \emph{disturbances}. Besides the
dynamics of the system, there may be limits on possible actions and states. That is, the models are only valid in a certain operating area.
These operating constraints may be due to technical limitations, regulations, safety measures, etc. System models may change over time. That is, the structural
parameters of the model need not stay constant.

We can distinguish four different types of system models: \emph{centralized},
\emph{decentralized}$\backslash$\emph{distributed}, and \emph{hierarchical} models.
\begin{itemize}
\item We may model the system with \emph{one system model}, describing the
whole system. This model may be very large if a high degree of detail is
required, or very abstract if this is not the case. We call such a model a
\emph{centralized model}. E.g., if we consider the system of a car, we can determine one single system model which describes the dynamics of the car completely.

\item In some cases, the overall system can naturally be seen as a
\emph{collection of smaller subsystems} that are completely decoupled from one another, or of which it is assumed that they are completely decoupled. Each system is autonomous. We refer to a system model consisting of several smaller
decoupled subsystem models as a \emph{decentralized model}. E.g., when we have a number of cars, the individual dynamics are decoupled and we have a decentralized setting. If we do have couplings between the subsystems, we have a \emph{distributed model}. E.g., a car that has another car connected to it with a rope can be modeled as a distributed system.

\item We may also be able to distinguish system models with \emph{different
layers of abstraction}. The highest layer may model the dominant characteristics
of the system, whereas lower layers may model more detailed characteristics.
Information at higher layers is typically used in lower layers and vice versa.
\end{itemize}
We can see a centralized model as a special form of a
hierarchical model in which there is only one layer and one system model.
Also a decentralized model can be seen as a hierarchical model in which
there is one layer with all the subsystems of the decentralized model and no
higher layer. And finally a distributed model can be seen as a hierarchical model by defining a two-level hierarchy in which the lowest
layer consists of the two subsystems, with links to a higher layer that connects
the variables of one system to the other.

\subsection{Control Problem Models}\label{sec:dcp}
Depending on the structure of the system model the overall goal may consist of
one \emph{centralized goal}, a set of \emph{decentralized goals}, or a set of
\emph{hierarchical goals}. When using a centralized goal there is one overall goal for the whole system. Decentralized goals appear when subsystems in an overall system each have their own independent goals. Hierarchical goals arise when subsystems have goals that (partially) overlap, or when goals for a system can be defined on different levels of abstraction/detail. The goals typically have a
close relation to (part of) the overall system.

Similar to the three different types of system models,
we can define three types of \emph{control problem models}:
\begin{itemize}
\item A \emph{centralized problem model} consists of one single DCP.
\item A \emph{decentralized problem model} consists of multiple smaller,
independent, DCPs. If the smalled DCPs have no conflicting goals, the combination of the problems is equivalent to the overall DCP. However, if there are conflicting goals, the combination of the problems need not be equivalent.
\item A \emph{hierarchical problem model} consists of a number of abstraction
layers, in which higher layers contain more abstract DCPs, and lower layers more
concrete DCPs. The higher layers depend on information from lower layers and
vice versa.
\end{itemize}
The structure of the problem model may be closely related to the structure of
the system model. However, this need not always be the case. E.g., we
may have a centralized system model with a hierarchical problem model, or vice
versa.

\subsection{Agent Architectures}
Solving DCPs is done through the use of \emph{controllers}, or \emph{agents}.
In general, agents are problem solvers that have abilities to act, sense,
reason, learn, and communicate with each other in order to solve a given
problem. Agents have an \emph{information set} containing their knowledge
(including information from sensing and communicating), and an \emph{action set}
containing their skills.

\begin{figure}
\centering \includegraphics[width=0.6\textwidth]{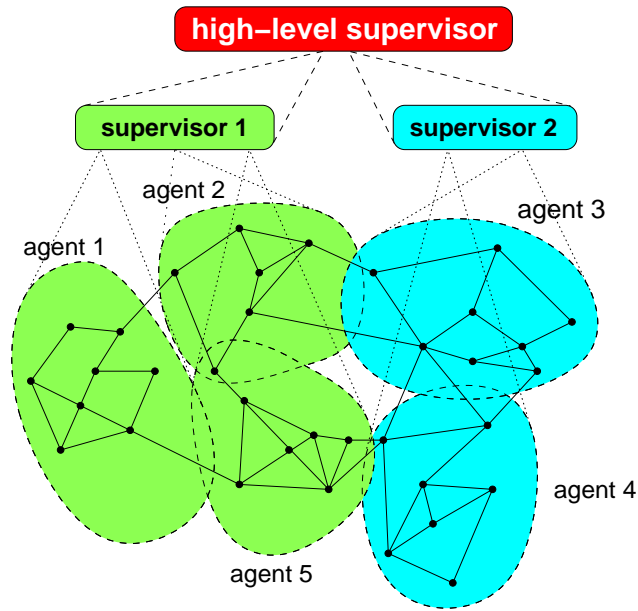}
\caption{Hierarchically structured agent architecture. Higher level agents provide targets for lower level agents. The agents at the lowest level control the physical system. Besides structured in layers, agents may also be organized in groups within which the agents are able to communicate with one another directly.
}
\label{fig:hierarchy}
\end{figure}

Agents may be organized in architectures, e.g., through communication
links. We can again distinguish three agent architectures:
\begin{itemize}
\item a \emph{centralized} agent architecture, in which there is only one
single agent,
\item a \emph{decentralized} agent architecture, in which there are
numerous agents that do not have any interaction among one another,
\item a \emph{hierarchical} agent architecture, in which there are
different layers of agents. Higher layers may supervise and receive
information from lower layers. Lower layers may follow instructions from and
provide information to higher layers. Agents on the same layer may be allowed to communicate directly with one another, or through the higher layers. See Figure \ref{fig:hierarchy} for an example of a hierarchically structured agent architecture.
\end{itemize}
Note that communication between two agents on the same layer can be replaced
by a virtual communication agent one layer higher in order to satisfy a
no-com\-mun\-ica\-tion-on-a-layer assumption. Note that when considering
hierarchical architectures, it is not only important to determine which
information is communicable, but also in which order information is accessible
to agents. That is, there needs to be a \emph{communication protocol}.

\subsection{Design Decisions}\label{sec:dd}
The models introduced in the current section leave many questions when
it comes to designing control systems. First of all, how should the system be
modeled? Is there a logical subsystem structure? Can it be found from a
centralized system model? Second, how should control problems be formulated on
the chosen system model. Third, how should the agent architecture be designed to
solve the control problem? More precisely, what acting, sensing, and
communication skills should agents have in order to solve the problems?
According to what protocol should they communicate with each other?

Sometimes the agent architecture already exists; in that case the questions
may be reversed: what subproblems can the agents solve? How can the subproblems
be designed in such a way that the overall goal of the system is obtained? How
should the subproblems be assigned to the agents?

At a higher level we may consider agents clustered in groups. How can agents be
clustered in groups such that the information exchanged within the group is
maximized and between groups minimized? Similarly, how can the agents
be clustered such that the combined skills in each group are sufficient to solve
the combined subproblems of the group?


\section{Single-Agent Model Predictive Control}\label{sec:sampc}
Over the last decades MPC \cite{CamBor:95,GarLee:00-doesnotexist,Mac:02} has
become the advanced control technology of choice for controlling complex, dynamic systems, in particular in the precess industry. In
this section we introduce the MPC framework. We relate the standard MPC
formulation to the models introduced in the previous section and find that MPC
can be referred to as single-agent MPC.

\paragraph{Control Design Characteristics}
\begin{figure}
\centering \includegraphics[width=0.7\textwidth]{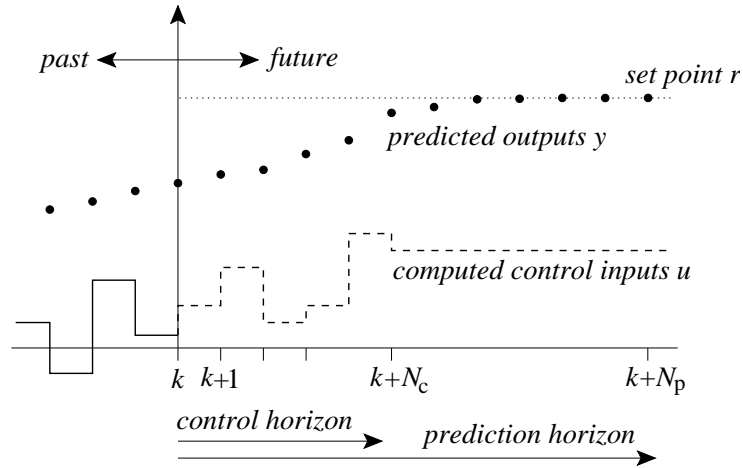}
\caption{Example of conventional MPC. The control problem is to find actions
 $u_k$ to $u_{k+N_\mathrm{c}}$, such that after $N_\mathrm{p}$ steps the system
behavior $y$ approaches the desired behavior $y^*$.
In this example, $y$ indeed reaches the desired set point $y^*$.
}
\label{fig:mpc}
\end{figure}
In terms of the previous section, the MPC
formulation is based on a \emph{centralized system model}, with a
\emph{centralized control problem model}, and a \emph{centralized agent
architecture}.
\begin{itemize}
\item The \emph{centralized system model} is given by a (possibly time-varying)
dynamic system of difference or differential equations and constraints on
inputs, states, and outputs.

\item The goal of the control problem is to minimize a
cost function. The control problem is stated as a single-objective
optimization problem.

\item The problem is solved by a single \emph{centralized agent}, the
information set of which consists of measurements
of the physical system, and the action set of which consists of all possible
 actions. The agent solves the problem with a three-step
procedure, see also Figure \ref{fig:mpc}:
\begin{enumerate}
\item It reformulates the control problem of controlling the \emph{time-varying} dynamic system using a \emph{time-invariant approximation} of the
system, with a \emph{control} and a \emph{prediction horizon} to make
finding the solution tractable, and a \emph{rolling horizon} for
robustness.

\item It solves the reformulated control problems, often using general,
numerical solutions techniques, while taking into account constraints on actions and
states.

\item It combines the solutions to the approximations to obtain a
solution to the overall problem. This typically involves implementing the
actions found from the beginning of the time horizon of the current
approximation, until the beginning of the next approximation.
\end{enumerate}
\end{itemize}

Since the MPC framework uses a single agent, we can refer to it as
\emph{single-agent MPC}.

\paragraph{Advantages}
Single-agent MPC has found wide success in many different applications, mainly
 in the
process industry. A number of advantages make the use of single-agent MPC
attractive:
\begin{itemize}
\item The framework handles \emph{input, state, and output constraints} explicitly
in a systematic way. This is due to the control problem formulation being based
on the system model which includes the constraints.

\item It can \emph{operate without intervention} for long periods. This is
due to the rolling horizon principle, which makes that the agent looks ahead to
prevent the system from going in the wrong direction.

\item It \emph{adapts easily} to new contexts due to the rolling horizon.
\end{itemize}

\paragraph{Disadvantages}
However, the use of single-agent MPC also has some significant
disadvantages:
\begin{itemize}
\item The approximation of the DCP with static problems can be of \emph{large
size}. In particular, when the control horizon over which actions are computed
becomes larger, the number of variables of which the agent has to find the
value increases quickly.

\item The \emph{resources} needed for computation and
memory may be high, increasing more when the time horizon increases. The amount
of resources required also grows with increasing system complexity.

\item The \emph{feasibility} of the solution to DCP is not guaranteed. Solutions
to the approximations do not guarantee solutions to the original DCP.
\end{itemize}

Research in the past has addressed these issues, resulting in conditions for
feasibility and stability, e.g., using contracting constraints, constraint relaxation, and
classical stabilizing controllers at the end of the horizon. Most of the MPC
research has focused on centralized computations. In the
following section we look at research directed at extending the single-agent MPC
 framework
to the use of multiple agents. Using multiple agents to tackle the control
problem may reduce the computational requirements compared to a single agent
approach.


\section{Multi-Agent Model Predictive Control}\label{sec:mampc}

In the remainder of this report we discuss the use of Multi-Agent MPC. As
the name suggests, in multi-agent MPC multiple agents try to solve the DCP.
 Although not
strictly necessary, when considering multiple problem solvers, it is typical to
have multiple different problems. The DCP is therefore typically broken up into
a number of smaller problems. The main advantage of this is that the computational
burden can be lowered. Agents can communicate and collaborate with other agents to come up
with a good solution. If the agents can work asynchronously then they can run in
parallel and at their own speed. This is a desirable situation for control of
large-scale systems. However, synchronization problems may be hard to solve.

Many authors have considered using MPC as part of a distributed
control architecture. Some examples of these are architectures in which a
single MPC controller is used as replacement of decentralized PID controllers (Pomerlea
\etal\cite{Pom:03}), multiple different MPC controllers are manually engineered
as replacement of decentralized PID controllers (Irizarry-Rivera \etal
\cite{Iri:97}, Ochs \etal\cite{Och:98}), or MPC is used as supervisory layer in
a cascaded setting (Silva \etal\cite{Sil:97}, Vargas-Villamil \etal
\cite{Var:00}). The control architectures involved are typically engineered with
insight in the specific application domain. Other architectures consider
multiple subsystems that depend on one another, and that employ MPC in order to
optimize system performance. These kind of applications are among others
discussed by Braun \etal\cite{Bra:03}, Katebi and Johnson \cite{Kat:97},
Georges \cite{Geo:94}, Camponogara \cite{Cam:02}, Aicardi \etal\cite{Aic:92},
Acar \cite{Aca:92}, Sawadogo \etal\cite{Saw:98}, El Fawal \etal\cite{Elf:98},
G\'omez \etal\cite{Gom:98}, Baglietto \etal\cite{Bag:99}, Jia and Krogh
\cite{Jia:01, Jia:02}, and Dunbar and Murray \cite{Dun:02}. In this survey we
mainly focus on this last class, since the methods described in this class are
more general (less application specific) and therefore more widely applicable than
the methods described in the first class.

\paragraph{Control Design Characteristics}
In general, the main difference between the multi-agent MPC and single-agent MPC
 framework is that in
the multi-agent MPC several agents are used to solve the DCP. We can
 characterize the
multi-agent MPC framework as follows:
\begin{itemize}
\item The system model is typically a \emph{hierarchical system model}.

\item The control problem is typically formulated to minimize a
\emph{hierarchical} cost function.

\item The control problems are typically solved by a \emph{hierarchical agent
architecture}.
\end{itemize}
In multi-agent MPC, the centralized system and control problem are first
\emph{decomposed} into smaller subproblems. The subproblems will in
general depend on each other. To solve the problems the agents therefore need
to \emph{communicate} with each other.

In this section we survey some of the approaches recent research has taken for
decomposing the DCP into sub-DCPs and finding a suitable solution to those. We
are particularly interested in seeing how different authors decompose the
overall system into subsystems, how they define the subproblems, and how agents
communicate with each other to come to a solution.

\subsection{System Model Decomposition}

Typically there are two ways in which a decentralized or hierarchical
system model is formed, based on the way the overall system is considered:
\begin{itemize}
\item The centralized system model can be used \emph{explicitly}. In this
case, a \emph{centralized} system model is first \emph{explicitly}
constructed and then decomposed into several subsystems using
structural properties that are present in the system model. This is a \emph{top-down} approach. E.g., Motee
and Sayyar-Rodsari \cite{Mot:03}, and Katebi and Johnson \cite{Kat:97}
analytically decompose a linear dynamic system into an equivalent set of
subsystems with coupled inputs.

\item The centralized system model can also only be considered
\emph{implicitly}. This means that the decomposition into subsystems is based on
engineering insight and typically involves modeling a subsystem and the
relations with other subsystems directly. In this case we have a \emph{bottom-up} approach. E.g., Georges \cite{Geo:94}, El
Fawal \etal\cite{Elf:98}, Braun \etal\cite{Bra:03}, and G\'omez \etal
\cite{Gom:98} design the subsystems without first considering a model for the
overall system. \end{itemize}
In general, as Sandell \etal\cite{San:78} point out in their survey,
dividing a system into subsystems may be done by considering different time
scales in a system and looking for weak couplings between subsystems. By our knowledge there are no generic methods to do the decomposition.

\paragraph{Decentralized Model Decomposition}
In our definition of a decentralized decomposition, all subsystems are
independent of one another. This situation is not discussed in the articles
surveyed for this report. However, many authors do use the word
\emph{decentralized} to address a group of subsystems that can communicate with
one another. We see this group of subsystems as a special case of a hierarchical
system model. Purely decentralized model decomposition only is possibly when two
subsystems are completely independent of each other, or when they are
\emph{assumed to be} independent of each other. The term \emph{decentralized} should not be confused with the more general term \emph{distributed}. The latter refers to systems consisting of subsystems in general, and not in particular to systems consisting of strictly independent subsystems.

\paragraph{Hierarchical Model Decomposition}
Hierarchical system model decomposition arises when subsystems depend on each
other, they are \emph{coupled}. Higher levels in a hierarchy may be more
\emph{abstract} or may span a \emph{longer time period} (e.g., they may have
a lower communication, computation, or control rate). The coupling between
subsystems can have different foundations:
\begin{itemize}
\item Sometimes the
coupling is based on \emph{physical variables} and modeled explicitly, like in
Sawadogo \etal\cite{Saw:98}. They consider control of a water system divided in
different sections as subsystems. In each subsystem model the controls and state
of a neighboring subsystem are taken into account. Dunbar and Murray
\cite{Dun:04} consider multi-vehicle formation stabilization. The
system models of the vehicles (including constraints) are uncoupled.
However, one layer higher, at a more abstract level, the state vectors of the
subsystems are coupled due to constraints that make the vehicles drive in a formation. Baglietto \etal\cite{Bag:99} consider optimal dynamic
routing of messages in a store-and-forward packet switching network. The nodes
in the network are seen as subsystems with connections to neighboring
subsystems. In particular, by reformulating the subsystem model they get rid of
constraints.

\item Sometimes the coupling is more \emph{artificial} and does not have a clear
physical meaning. E.g., Georges \cite{Geo:94} and El Fawal
 \etal\cite{Elf:98} define a subsystem model for each section in a water
distribution network. They introduce \emph{compatibility} equations between
subsystems that have to be satisfied. The decentralized approach is based on an
augmented Lagrangian formulation, where the flow balancing equations are
dualized. In this formulation, the Lagrangian multipliers become the coupling
variables.
\end{itemize}

\subsection{Control Problem Decomposition}

In the reviewed literature there is no distinction between the structure
of the system model and the structure of the problem. So, with each
subsystem a control problem is associated with its own goal. We believe that
in general it may not always be necessary to assign a control problem with
each subsystem. It may be easier and sufficient to define goals over a number of
subsystems, rather than for each subsystem individually. In the reviewed
literature, the goals of the subproblems are obtained from:
\begin{itemize}
\item an \emph{analytical
decomposition} of a centralized goal. E.g., Georges \cite{Geo:94} and El
Fawal \cite{Elf:98} take some sort of worst-case approach by defining for each subsystem a subproblem of finding the
Lagrangian multipliers that maximize the problem of finding the controls that
minimize the augmented Lagrangian.

\item an \emph{ad-hoc engineered} subproblem goal. E.g., Baglietto \etal
\cite{Bag:99} formulate a goal for each subsystem. \end{itemize}
Dunbar and Murray
\cite{Dun:04} consider no special goal for the lowest layer. However, one layer
higher, a centralized goal is defined over the subsystems of the lowest level.
Katabi and Johnson \cite{Kat:97} take a similar approach. Jia and Krogh
\cite{Jia:02,Jia:01} consider agents that exchange predictions on the bounds of
their state trajectories. Thus agents have information about the trajectories
that the subsystem of the other agents will potentially make.

\subsection{Problem Solving}

As mentioned in Section \ref{sec:dd}, it is important to consider the problem of
designing the agent architecture and assigning suitable subproblems to the
agents in the architecture. Once this assignment has been made a suitable
coordination protocol should be used in order to find a good solution. This coordination protocol specifies how and what information is exchanged between agents.

\subsubsection{Agent Design and Problem Assignment}

\paragraph{Agent Design}
The information set of the agents is often implicitly assumed to contain
sufficient information for solving the subproblems. Also the action sets of
the agents are assumed to be sufficient (Georges \cite{Geo:94},
El Fawal \cite{Elf:98}, Baglietto \etal\cite{Bag:99}). If an agent does not
have access to certain information that it needs directly it has two options:
obtain the information through \emph{communication}, or have some means to
\emph{predict} the information.

There has been some interesting research put into optimally
partitioning agents into groups \cite{Cam:02}. Motee and Sayyar-Rodsari
\cite{Mot:03} remark that the elimination of the communication
requirements between agents (at least among agents in different sub-groups) is
of crucial importance. The less communication between agents, the easier
they can work at their own speeds. This requirement must be balanced against the
total cost of control actions. The authors propose a formulation in which such a
trade-off can be trivially exercised by finding a matrix assigning
agents to groups.

Motee and Sayyar-Rodsari \cite{Mot:03} also consider how information must be
communicated to the groups. They propose a sensitivity-based criterion. For the
system with the grouped agents, the sensitivity of the closed-loop control
action to the output measurement can be used as a criterion for deciding whether
a certain output measurement must be made available to that group of agents.
This analysis is done offline.

\paragraph{Problem Assignment}
In the reviewed literature each control subproblem is assigned a specific
agent to solve the problem. Most designs for agent architectures are
made offline and do not change online \cite{Kat:97}. The information that agents
may share with each other is determined a priori by, e.g., minimizing a minimal communication cost, or objective at the level of the overall
system.

\subsubsection{Coordination Schemes}

The way in which agents communicate with one another is crucial in whether or not a useful, feasible, preferably optimal, solution is obtained. Agents communicate and exchange information according to a certain \emph{coordination scheme}. Important attributes of these schemes are: \emph{iterative solution}, \emph{choice of actions}, \emph{subproblem modification}, and \emph{automatic learning}
\cite{Cam:02}. Choices for these attributes have directly influence the performance of solving the control problem. In the following we will go into more detail on these items and relate them to the existing literature.

\paragraph{Iterative Solutions}
\begin{figure}
\centering \includegraphics[width=0.65\textwidth]{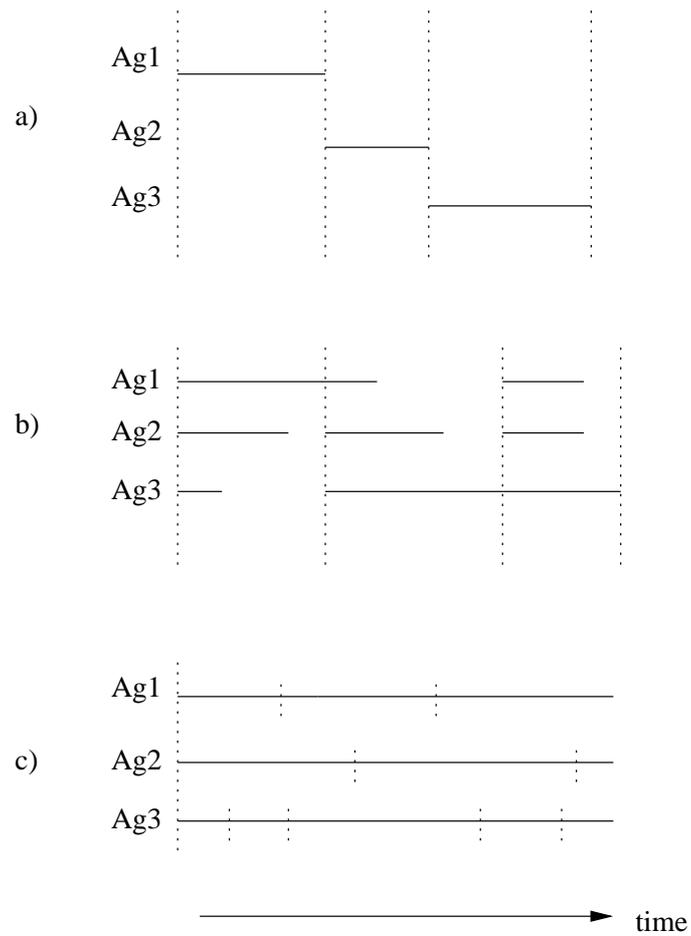}
\caption{Different types of iterative schemes. a) Serial, synchronous: one agent takes a step at a time, after which a next agent takes a next step; b) Parallel, synchronous: all agents take a step at a time, but they wait with taking the next step until all agents are finished taking the step of the current time; c) Asynchronous: all agents take steps at their own speed and they do not wait for one another.
}
\label{fig:iter}
\end{figure}
It is often convenient and practical to find a solution by iterations,
particularly when decisions are shared among agents whose goals are
conflicting \cite{Cam:02}. Each agent continually revises its decisions, taking
into account the decisions of its neighbors. If the dynamic subproblems are
decoupled, then the agents reach optimal decisions independently. If this is not
the case, the couplings can be dealt with in two ways: \emph{synchronously} or
\emph{asynchronously}. See also Figure \ref{fig:iter}.
\begin{itemize}
\item In the
\emph{synchronous} case precedence constraints are imposed on the iterations,
which makes that faster agents have to wait for slower agents. A distinction can
be made between \emph{serial} synchronous methods and \emph{parallel}
synchronous methods. In the former case only one agent takes a step at a time.
In the latter agents wait for all other agents to finish the current step before
proceeding.

\item \emph{Asynchronous} treatment allows all agents to run at their own speed
and is therefore preferred over the synchronous case, since the agents spend no time waiting for one another. However, this comes at the price of uncertainty in information, since agents might not know exactly what other agents will do.
\end{itemize}

Georges \cite{Geo:94} and El Fawal \etal\cite{Elf:98} deal
with parallel synchronization in their two-step algorithm. In their approach
first each agent solves its subproblem using certain parameters of the
previous step. These parameters are optimized using information from other agents. This information is obtained in the second step in which each agent communicates its parameters with the other agents. In some cases agents communicate their expected plans to each other
after each optimization step. For example, Jia and Krogh \cite{Jia:01,Jia:02}
let the agents solve local min-max problems to optimize performance with respect
to worst-case disturbances. Parameterized state feedback is introduced into the
multi-agent MPC formulation to obtain less conservative solutions and predictions.
Dunbar and Murray \cite{Dun:04} have a similar approach. Shim \etal\cite{Shi:03}
also include the capability for agents to combine the trajectory generation
with operational constraints and stabilization of vehicle dynamics by adding to the cost function a potential function reflecting the state information of a possibly moving
obstacle or other agent.

\paragraph{Choice of Action}
Agents can apply different ways of actually choosing which action to perform at
a certain point. Typically, in single-agent MPC, the agent implements the first action of
the action sequence found by solving its control problem. However, in multi-agent MPC
there are alternatives since an agent may accept \emph{suggestions} from
neighboring agents regarding its actions. These suggestions could be, e.g., values to set:
\begin{itemize}
\item An agent may \emph{exclusively} choose its action and implement
it, e.g., by simply choosing the first action taken from the
sequence of optimal actions. Georges \cite{Geo:94} and El Fawal \etal
\cite{Elf:98} consider a higher-layer agent that obtains information from
multiple lower layers to exclusively update parameters for each agent.

\item Actions may be \emph{shared}. That is, actions of a certain agent can be
chosen by other agents as well. In \cite{Cam:00}, Camponogara
discusses this. Other agents may be allowed to use capabilities of a certain agent that only that agent has.  

\item Agents may choose which action to perform in a \emph{democratic} way by
letting multiple agents vote about which action to take.
Camponogara \cite{Cam:00} shows that this can be beneficial, assuming that the majority knows what is right to do.

\item Agents may \emph{trade} their actions, in which
case the agent with the highest bid gets to choose the action that an agent will
perform. This might be useful in situations where there is a limited resource that needs to be shared.
\end{itemize}

\paragraph{Subproblem Modifications}
Ideally protocols that agents use for cooperation can deal with the
subproblems of the agents directly. Sometimes a protocol may however to some
extent require the modification of the subproblems. It may demand\begin{itemize}
\item the reformulation of subproblems as \emph{unconstrained}
subproblems, that is, to remove all limits,
\item the \emph{relaxation} of subproblems with tolerance factors, that is, to allow going over certain limits,
\item \emph{tightening} of subproblems with
resource factors, that is, to lower the limits as much as possible.
\end{itemize}
In particular when asynchronously working agents are considered, these
modifications may be needed. As an example, each agent needs to know what the
other agents might want to do, so it can anticipate these actions if it
chooses to be unselfish. Shared resources need to be shared in ways that seem
fair. However, faster agents may grab all the resources. Resource factors may
help here. Camponogara \cite{Cam:00} investigates each of the
modifications. Jia and Krogh
\cite{Jia:01,Jia:02} impose predicted state bounds as constraints in
subsequent multi-agent MPC iterations to guarantee their subsystem satisfies the bounds
broad-casted to other agents.

\paragraph{Automatic Learning}
Automatic learning may boost the effectiveness, widen the scope of applications
and improve the adaptability of cooperation protocols. Learning can for
example be introduced for \emph{parameter identification}, or for
\emph{improvement of problem-solving} and \emph{decision-making} abilities.
Learning may enable an agent to predict what its neighbors will do. Learning is
in particular useful when agents work asynchronously.

Georges \cite{Geo:94} and Katabi \etal\cite{Kat:97} include an on-line
identification procedure based both on a MIMO parametrized model of the physical
characteristics of the system and Kalman filtering. Besides that
the authors use a Kalman optimal estimator defined on the basis of the on-line
identified control to estimate the state of the subsystems over which the
subproblem of an agent is defined. Baglietto \etal\cite{Bag:99} assign neural
networks to the agents representing nodes in a network. These neural networks
are trained offline to improve online computational requirements. In G\'omez
\etal\cite{Gom:98}, the models of the nodes depend on future state values of
neighboring nodes. Each agent estimates these values.


\subsection{Conditions for Convergence}

Motee and Sayyar-Rodsari \cite{Mot:03} remark that the optimal action for a
subproblem can only be obtained if the optimal action to the other
subproblems is available (when the problems depend on each other). This is also
discussed by Talukdar \etal\cite{Tal:00} with elements from game theory.

Let the \emph{reaction set} of an agent contain the actions that
the agent would make when it knows what the other agents will do. The set of
\emph{Nash equilibria} is the intersection of the reaction sets of all the
agents. The \emph{Pareto set} is the set of feasible solutions to the overall
problem. Talukdar \etal\cite{Tal:00} make three observations:

\begin{itemize}
\item The elements of the Pareto set are the best trade-offs among the
multiple objectives of the subproblems. These may be better trade-offs
than those provided by Nash equilibria.

\item Constraints can change the solution sets of (sub)problems significantly.
With techniques like Lagrange multipliers, penalty functions and barrier
functions, it is always possible to convert a constrained problem into an
unconstrained one. However, these conversions should be used with care. Both
conceptually and computationally it is advantageous to preserve the separate
identities of constraints, not the least of which is the option of specialized,
adaptive handling of each constraint during the solution process.

\item The solution to the overall centralized problem and the completely
decentralized problem are two extremes. The centralized problem is often
intractable but the Pareto solutions are the best that can be obtained. The
subproblems are smaller and more tractable. For an agent, the collection of the
solutions for all possible actions of its neighboring agents constitute
its reaction set. The intersection of the reaction sets of all agents identify
the Nash equilibria. The calculation of a reaction set requires the repeated
solution of the subproblems, which can be tedious.
\end{itemize}
Depending on the cooperation scheme used, the resulting performance will be
different. When considering multi-agent MPC it is important to look at the question
whether or not the agents are capable of cooperatively obtaining an optimal
solution to the overall control problem.

In \cite{Cam:02}, Camponogara \etal\ consider under what conditions iterations
converge to a solution of the subproblems and under what conditions the
solutions of the subproblems compose a solution to the overall problem.
Baglietto \etal\cite{Bag:99} remark that team-optimal control problems can be
solved analytically in very few cases, typically when the problem is LQG and the
information structure is partially nested, i.e., when any agent can reconstruct
the information owned by the decision makers whose actions influenced its
personal information. Aicardi \etal\cite{Aic:92} address the problem of the
existence of multi-agent MPC stationary control strategies in an LQG decentralized
setting. The possibility of applying a multi-agent MPC control scheme derives from the
assumptions on the information structure of the team. The authors of \cite{Aic:92} show how applications
of such a scheme generally yield time-varying control laws, and find a condition
for the existence of stationary multi-agent MPC strategies, which takes only a-priori
information about the problem into account. Dunbar and Murray \cite{Dun:04}
establish that the multi-agent MPC implementation of their distributed optimal control
problem is asymptotically stabilizing. The communication requirements between
subsystems with coupling in the cost function are that each subsystem obtains
the previous optimal control trajectory of the other subsystems to which it is coupled at each receding
horizon update. The authors of \cite{Dun:04} note that the key requirement for stability is that each
action sequence computed by the agents does not deviate too far from the sequence that has been computed and communicated previously.

Camponogara \etal\cite{Cam:02} develop conditions on the agents' problems and
cooperation protocols that ensure convergence to
optimal attractors. Unfortunately, they find that the conditions have some
severe disadvantages for practical use:
\begin{itemize}
\item The convexity of the overall objective function and constraints cannot be
guaranteed in practice. The dynamics of real-world networks can be highly
nonlinear and nonconvex.

\item The protocols developed in \cite{Cam:02} require that the initial solution is feasible. This feasibility is hard to meet. The
specifications on how the network should behave in the future can introduce
conflicts and make the problem infeasible. The resolution of the conflicts
stands as a hard problem.

\item The differentiability of the objective and constraint functions cannot be
expected in real-world problems. When the decisions are a mix of discrete and
continuous variables, non-differentiability is introduced in
the functions. The protocols developed in \cite{Cam:02} cannot deal with this.

\item The exact match of agents to subproblems is impractical. This means that each subproblem has a specific agent capable of obtaining the information and  making the actions needed to solve the subproblem. The agent
architectures could become too dense to induce an exact match.

\item The protocols developed in \cite{Cam:02} use interior-point methods. The use of these methods is not convenient since interior-point
methods are sensitive to implement and less robust than algorithms such as
sequential quadratic programming.

\item The enforcement of serial work within neighborhoods is quite impractical
and unattractive. The convergence speed would be very slow and therefore the network of agents would not respond promptly to disturbances.
\end{itemize}
Although these issues may be difficult to remove, Camponogara \etal\ show that it is
not always necessary to fulfill all the conditions. However, there are no general conditions under which this is not necessary.

\section{Conclusion}\label{sec:conc}

In this report we have given an overview of recent literature on multi-agent MPC. We have identified common aspects in each of the reviewed papers. This has led to
the identification of certain groups and attributes at a rather non-math\-e\-matical
level. This allows us to identify directions for further research. Although
since the survey paper of 1978 by Sandell \etal\cite{San:78} a significant amount of progress has been made, many issues remain
to be investigated. Some of these are:
\begin{itemize}
\item The decomposition of system models and control problems may be automated.
Methods could be developed that propose different decompositions.

\item In the current literature only one or two layers are considered.
It is not clear how hierarchies with more layers can be automatically used. Perhaps more layers can be used in similar ways or with some form of nesting.

\item The assignment of subproblems to agents may be automated. Perhaps agents
can negotiate about who solves which subproblem. The assignment should be efficient, robust, etc.

\item The conditions for convergence to optimal solutions to the
overall problem are too restrictive for practical application. Perhaps
classes of systems or control problems could be identified in which multi-agent MPC may have fewer conditions for
convergence.

\item Until these classes have been identified, heuristic coordination schemes
need to be developed that give good results. Further research into asynchronous
cooperation protocols is needed.
\end{itemize}
With enough research in these directions, applications of truly autonomous
multi-agent control may become possible.

\section*{Acknowledgments}
This research was supported by project ``Multi-agent control of large-scale
hybrid systems'' (DWV.6188) of the Dutch Technology Foundation STW, Applied
Science division of NWO and the Technology Programme of the Dutch Ministry of
Economic Affairs. 


\providecommand{\noopsort}[1]{}\providecommand{\sahinidummy}{}

\end{document}